\documentclass[amsmath,amssymb,nofootinbib]{revtex4}

\usepackage{latexsym,comment,verbatim}
\usepackage{amssymb}
\usepackage{amsfonts,hyperref}
\usepackage{amsmath,color}
\usepackage[draft]{graphicx}
\usepackage{bm}

\def\mb#1{\mathbf{#1}}

\def\ber{\begin{eqnarray}}
\def\eer{\end{eqnarray}}
\def\beq{\begin{equation}}
\def\eeq{\end{equation}}

\def\ed{\end{document}}
\def\di#1#2{\frac{\mathrm{d} #1}{\mathrm{d}#2}}

\begin{document}

\author{Angelo Tartaglia}
\email{angelo.tartaglia@inaf.it}
\affiliation{INAF-OATo}

\author{Matteo Luca Ruggiero}
%\email{matteo.ruggiero@polito.it}
\affiliation{Politecnico di Torino, Corso Duca degli Abruzzi 24, 10129 Torino - Italy \\ INFN, LNL - Legnaro, Italy}
\date{\today}

\title{From Kerr to Heisenberg}

\begin{abstract}
In this paper we consider the space-time of a charged mass endowed with an
angular momentum. The geometry is described by the exact Kerr-Newman
solution of the Einstein equations. The peculiar symmetry, though exact, is
usually described in terms of the gravito-magnetic field originated by the
angular momentum of the source. A typical product of this geometry is
represented by the generalized Sagnac effect. We write down the explicit
form for the right/left asymmetry of the times of flight of two
counter-rotating light beams along a circular trajectory. Letting the circle
shrink to the origin the asymmetry stays finite. Furthermore it becomes
independent both from the charge of the source (then its electromagnetic
field) and from Newton's constant: it is then associated only to the
symmetry produced by the gravitomagnetic field. When introducing, for the
source, the spin of a Fermion, the lowest limit of the Heisenberg
uncertainty formula for energy and time appears.
\end{abstract}

\maketitle

%------------------------Section-------------------------
\section{Introduction}
%------------------------Section-------------------------

The one century long story of the parallel development of the two main
physical theories of the 20th century is far from an end and continues to
puzzle our understanding with being both brilliantly successful in their
proper domain and in the same time diverging or even being in conflict in
the areas where they are obliged to cohabit. We are obviously referring to
General Relativity (GR) on one side and Quantum Mechanics (QM) on the other.
We will not enter here the intricate problem usually called "quantization of
gravity". An important portion of the theoretical physicists community
continues to work on it, mostly with an intellectual prejudice on the fact
that space-time \textit{must }be quantizable as if it were a field more or
less like others and so confusing the container (space-time) with the
content (fields).

We will simply try to evidence in the following a peculiar case where a link
is directly manifested between GR and QM, without the need to introduce new
tools and modifications of one or the other theory.

One of the weakest effects of the gravitational interaction as described by
GR has been dubbed \textit{gravitomagnetism}. It depends on the off-diagonal
time-space terms of the metric tensor, $g_{0i}$.\footnote{%
The usual conventions are applied for notation: indices running from $0$ to $3
$ are denoted by Greek letters; space indices running from $1$ to $3$ are
represented by Latin letters. The Einstein summation convention of equal co-
and contra-variant indices always holds.} We will then concentrate on the
travel times of test particles moving with locally constant velocity along
trajectories closed in space. "Closed in space" is not an absolute statement
since closedness in three space dimensions depends on the choice of the
reference frame; however there is something which does not depend on such
choice. When either the observer or the source of gravity or both is \textit{%
rotating }the times of travel of test particles moving with the same local
speed in opposite directions along the same closed trajectory (closed in the
reference frame of the given observer) turn out to be \textit{different}.
The simplest and typical case is the one known as the Sagnac effect \citep{Sagnac}: a
rotating observer, in flat (Minkowski) space-time, sending light signals
along a closed contour in opposite directions\footnote{%
Only excluding the case in which the contour is contained in a plane
parallel to the rotation axis.}  finds that the times
of flight are different even though the path is the same. The difference is
in terms of proper time of the observer, i.e. it is (proportional to) the
length of a space-time \textit{interval},\textit{\ }which means that it does
not depend on the reference frame and stays there for any observer \citep{TaRu2015}. In flat
space-time a rotating observer is of course not inertial and the closed path
of the test particles (be they light beams or locally isotachic objects)
must be obtained using appropriate physical devices. The time of flight
(tof) asymmetry may be described using a different narration by different
observers, but the result is exactly the same for everybody. In a rotating
reference frame we have non-zero off diagonal terms in the metric, whilst in the
frame of an inertial observer the metric is diagonal: the asymmetry is the
same for both. This is however Special Relativity: we are now interested in
GR.

Our case considers the space-time of a spinning source of gravity i.e., in
geometrical terms, of curvature. Now we have in general to do with a metric
with non-null $g_{0i}$ terms. It is always true that we may locally
diagonalize the metric (this happens because, apart from singularities,
there always is a local tangent flat space-time); the diagonalization is
however not contextually possible everywhere. The nickname
"gravitomagnetism" (GM) is, as known, due to the fact that, in weak field
approximation, the Einstein equations in vacuo assume a form practically
coinciding (but for a factor of $2$) with the classical Maxwell equations of
electromagnetism (EM)\ and $g_{0i}$ may, in the same approximation, be
treated as the components of a three-dimensional vector potential in analogy
with EM \citep{TaRu2003,Ruggiero:2002hz,Mashhoon:2003ax,pfister2014gravitomagnetism,Ummarino:2017bvz,Ummarino:2021vwc,Ummarino:2020loo,Ruggiero_2020,Ruggiero:2020oxo}.

Here, however, we shall use no approximation though continuing to use the GM
terminology. In this framework we shall evidence the somehow surprising
connection between GR and QM.

%------------------------Section-------------------------
\section{Generalized Sagnac effect} \label{sec:sagnac}
%------------------------Section-------------------------

We consider the line-element in the form
\beq
ds^{2}=g_{00}c^{2}dt^{2}+2g_{0i}cdtdx^{i}+g_{ij}dx^{i}dx^{j} \label{eq:metricastazionaria}
\eeq
where $g_{\mu\nu}=g_{\mu\nu}(\mb x)$, hence the space-time metric does not depend on time. The above metric is quite general in its form and, in particular,  it is said to be  \textit{non time-orthogonal}, because $g_{0i} \neq 0$. These off-diagonal terms are related to the rotation of the sources of the gravitational field and  on the rotational features of the reference frame: in the case of a rotating frame in flat space-time, they depend on the rotation rate; more in general, they express the rotation rate of the frame with respect to a Fermi-Walker tetrad (see e.g. \cite{rindler1990rotating}).

We want to calculate the asymmetry in the propagation times of two signals  in the space-time described by the line-element (\ref{eq:metricastazionaria}); this asymmetry is the so-called \textit{Sagnac time delay} (see e.g. \cite{rizzi2004relativistic,Ruggiero:2014hfa}). 
We consider two messengers (massive or massless particles) simultaneously emitted at a given location:  they propagate in opposite directions along the same path and reach the emission point at different times, thus evidencing an asymmetry.
In doing so, we need to impose some conditions to say that the particles propagating in the two opposite directions are identical but differ only for the direction of propagation: this is naively related to their speed. However, the coordinate speed $\displaystyle w^{i}=\di{x^{i}}{t}$ has not a direct physical meaning. If we want to give an operational meaning (i.e. in terms of observable quantities) to the speed of a particle, we may proceed as follows. Let us consider a set of observers located along the given closed trajectory and mutually at rest. One of them is the main observer $O$ from which the travel of the oppositely moving particles starts. For any observer of the group 
(let us call it $O1$) it is possible to introduce an inertial frame, relative to which
$O1$ is at rest: this is the so-called \textit{Locally Co-Moving Inertial Frame} (LCIF). In this frame, the proper
element of distance $d\sigma$ and time $dT$ can be defined in terms of the metric elements and coordinates
intervals and the speed of the particle when passing by $O1$ is well defined. It is (see \cite{landau2013classical,rizzi2004relativistic})
\beq
d\sigma^{}=\sqrt{\gamma_{ij}dx^{i}dx^{j}},\quad dT^{}=-\frac 1 c \frac{g_{\mu 0}}{\sqrt{-g_{00}}}dx^{\mu}
\eeq
where $\displaystyle \gamma_{ij}=\left(g_{ij}-\frac{g_{i0}g_{j0}}{g_{00}} \right)$. If we use these expressions the line-element (\ref{eq:metricastazionaria})  takes the Minkowskian form
\beq
ds^{2}=d\sigma^{2}-c^{2}dT^{2} \label{eq:metricadsigmadT}
\eeq

Consequently, for an observer at rest in this LCIF the speed of a moving particle is  $v=\di{\sigma}{T}$, that is the ratio between the proper element of distance $d\sigma$ traveled in a proper time interval $dT$.  This speed is well defined from an operational viewpoint and, as we are going to show, it is useful  to define a natural condition on the properties of the two counter propagating particles.

On substituting in (\ref{eq:metricadsigmadT}), we get
\beq
ds^{2}= \left (1-\frac{c^{2}}{v^{2}} \right) d\sigma^{2} =  \left (1-\frac{c^{2}}{v^{2}} \right)\gamma_{ij} dx^{i}dx^{j}
\eeq
and from  (\ref{eq:metricastazionaria}) we obtain
\beq
 \left (1-\frac{c^{2}}{v^{2}} \right)  \gamma_{ij}dx^{i}dx^{j}= g_{00}c^{2}dt^{2}+2g_{0i}c dt dx^{i}+g_{ij}dx^{i}dx^{j}
 \label{eq:dt1}
\eeq

Eq. (\ref{eq:dt1}) can be solved for the coordinate  time interval $dt$; to this end, we introduce $\displaystyle \beta \doteq v/c$. Notice that for light-like particles, on setting $ds^{2}=0$, we get  $\beta=1$, in agreement with the second postulate of special relativity,  and the left hand side of Equation (\ref{eq:dt1}) is equal to zero.  On using the definition of $\gamma_{ij}$, Eq. (\ref{eq:dt1}) now reads
\beq
0 = g_{00}c^{2}dt^{2}+2g_{0i}c dt dx^{i}+ \left(\frac{1}{\beta^{2}} \gamma_{ij}+\frac{g_{i0}g_{j0}}{g_{00}}\right)dx^{i}dx^{j}
 \label{eq:dt11}
\eeq
from which we obtain the two solutions
\beq
dt_{\pm}= \frac{1}{|g_{00}|c} \left(g_{0i}dx^{i} \pm \frac{1}{\beta}\sqrt{|g_{00}|\gamma_{ij}dx^{i}dx^{j}} \right)
= \frac{1}{|g_{00}|c} \left(g_{0i}dx^{i} \pm \sqrt{|g_{00}|} \frac{|d\sigma|}{\beta} \right) \label{eq:dtsol1}
\eeq
In the above result, we may distinguish two contributions to the propagation time: the first is due to the synchronization of distant clocks (see e.g. \cite{landau2013classical}) in the metric (\ref{eq:metricastazionaria}); the second describes the time occurring to cover the (proper) distance $|d\sigma|$ with the local speed $v=\beta\, c$. In particular, we see that the second contribution in (\ref{eq:dtsol1}) does not depend on the propagation direction, since it does not change when $dx^{i}\rightarrow -dx^{i}$ if we assume that \textit{ the speed $v$ (or equivalently $\beta$) is a function only of the position along the path}; the case $v=\mathrm{constant}$ along the path is a particular sub-case. This amounts to saying  that, in any LCIF along the path, particles  have the same velocity $v$ in opposite directions.  On the contrary, the other contribution depends on the propagation direction. 
Once the propagation path is known, (\ref{eq:dtsol1}) can be integrated  to obtain the coordinate time interval (remember that we are interested in the future oriented branch of the light cone). Then, the asymmetry in the propagation times is given by 

\beq
\Delta t = \frac 2 c \oint_{\ell} \frac{g_{0i}dx^{i}}{|g_{00}|}=-\frac 2 c \oint_{\ell} \frac{g_{0i}dx^{i}}{g_{00}} \label{eq:formulafond}
\eeq

Of course the particles take different times for propagating along the path, depending on their speed,  but what we have just shown is that
\textit{ the
difference between these times is always given by eq. (\ref{eq:formulafond}), in any stationary space-time, and for arbitrary paths,} both for matter and light particles, independently of their physical nature.

%------------------------Section-------------------------
\section{Geometric approach to the gravitomagnetic clock effect}
%------------------------Section-------------------------
   A variant of the generalized Sagnac effect is the so called \textit{GM} \textit{clock effect} \citep{gclock}: two identical clocks freely falling along a spatially closed orbit, in opposite directions, show a growing synchronization defect each time they cross each other. Whenever the orbit is circular (then the motion is uniform) the clock effect \textit{\ }may also be
described in terms of Minkowski geometry on the flat $1+1$ dimensional
surface of a cylinder (see \citet{tartaglia2000}).

    Suppose that two identical clocks move in opposite directions along the same circular orbit at radius $r=R$. At proper time $\tau =0$ they cross at some point of the orbit and turn out to be synchronous there. The orbital
    speeds are in general different; let us call $\omega _{+}$ the one of the corotating clock (same sense as the central mass) and $\omega _{-}$ the one of the counter-rotating clock; it is $\omega _{+}>0$ and $\omega _{-}<0$. The proper times shown by the two clocks, when they meet for the second time, will in general be different. Under the assumed conditions the equations of motion along the orbit are plainly linear and, as seen by a distant observer, may be written
    \begin{equation}
        \phi_+=\omega_+ t ; \qquad  \phi_-=\omega_- t
    \end{equation}
    The meeting point $C$ satisfies the obvious condition $\phi_{C+}-\phi_{C-}=2\pi$ at the same coordinate time $t=t_C$.
    It is:
    \begin{equation}
        t_C=\frac{2\pi}{\omega_+ - \omega_-}
    \end{equation}
    This condition, using Minkowski  geometry,\footnote{In practice the calculation is the one for the length of a side of a scalene triangle, given the other two and in  presence of a Lorentzian signature} leads to the proper times of the two clocks \citep{tartaglia2000}:
    
    \begin{equation}
\left\{
\begin{array}{c}
\tau _{C+}=t_{C}\sqrt{g_{00}+2g_{0\phi }\omega _{+}+g_{\phi \phi }\omega
_{+}^{2}}=\frac{2\pi }{\omega _{+}-\omega _{-}}\sqrt{g_{00}+2g_{0\phi
}\omega _{+}+g_{\phi \phi }\omega _{+}^{2}} \\
\tau _{C-}=t_{C}\sqrt{g_{00}+2g_{0\phi }\omega _{-}+g_{\phi \phi }\omega
_{-}^{2}}=\frac{2\pi }{\omega _{+}-\omega _{-}}\sqrt{g_{00}+2g_{0\phi
}\omega _{-}+g_{\phi \phi }\omega _{-}^{2}}%
\end{array}%
\right.   \label{TC}
\end{equation}
    
These results are exact and, as far as it is $\omega _{+}\neq \omega _{-}$ it is also $\tau _{C+}\neq \tau _{C-}$. In the case of freely orbiting test masses, the two orbital velocities appearing in the $\tau _{C\pm }$ may be
determined solving the equations for the geodesics of the considered space-time and imposing the conditions assumed for the problem.
    
All the above corresponds to the strict definition of the \textit{clock effect}; however another case, recalling the generalized Sagnac effect, is the difference of the tof for the two clocks coming back where observer $O$ is. Now the spanned angle is in any case $2\pi $ if we assume the local
observer to be at rest with the distant one. If the angular velocities in the two senses are different we expect also the periods $T_{+}$ and $T_{-}$ to be different:%
\[
T_{\pm }=\frac{2\pi }{\left\vert \omega _{\pm }\right\vert }
\]
Passing to the corresponding proper times of the two "messengers" (again using the geometrical approach) it is

\[
\tau _{O\pm }=\sqrt{g_{00}T_{\pm }^{2}\pm 4\frac{\pi }{c}g_{0\phi }T_{\pm }+4%
\frac{\pi ^{2}}{c^{2}}g_{\phi \phi }^{2}}
\]

The time asymmetry $\Delta \tau _{O}=\tau _{O+}-\tau _{O-}$ will be obtained
comparing the readings on two clocks when they come back to $O$, now at different times. In the case of light we
have to do with null world-lines and a difference is found looking at the
clock of the observer in $O$ (or considering the beat frequency in a ring
laser at rest with $O$ and coinciding with the closed contour travelled by
the light beams \citep{ginger}). The condition is now:

\[
g_{00}T^{2}+4\frac{\pi }{c}g_{0\phi }T+4\frac{\pi ^{2}}{c^{2}}g_{\phi \phi
}=0
\]%
whence

\[
T_{\pm }=\mp \frac{2\pi }{c}\frac{g_{0\phi }}{g_{00}}+\frac{2\pi }{c}\frac{%
\sqrt{g_{0\phi }^{2}-g_{00}g_{\phi \phi }}}{g_{00}}
\]

The tof asymmetry is:

\begin{equation}
\Delta T=T_{-}-T_{+}=\frac{4\pi }{c}\frac{g_{0\phi }}{g_{00}}
\label{eq:fondcirc}
\end{equation}%
coinciding in practice with (\ref{eq:formulafond}).

All this is in terms of global coordinates; if we wish to know what is
locally measured by observer $O$ we have to multiply by $\sqrt{g_{00}}$ and
we obtain

\[
\Delta T_{O}=\frac{4\pi }{c}\frac{g_{0\phi }}{\sqrt{g_{00}}}
\]

The situation is a bit more complicate if $O$ is moving, and of course it is
moving if it is in free orbital motion around the central mass. Assuming
again that the trajectory is a circle, let us call $\Omega $ the orbital
velocity with respect to the distant observer. Projecting $\Delta T_{O}$
onto the worldline of the moving observer yields:

\begin{equation}
\Delta T_{\Omega }=\Delta T\frac{\sqrt{g_{00}}+\frac{g_{0\phi }}{\sqrt{g_{00}%
}}\frac{\Omega }{c}}{\sqrt{g_{00}+2g_{0\phi }\frac{\Omega }{c}+g_{\phi \phi }%
\frac{\Omega ^{2}}{c^{2}}}}=\frac{4\pi }{c}\frac{g_{0\phi }}{g_{00}}\frac{%
\sqrt{g_{00}}+\frac{g_{0\phi }}{\sqrt{g_{00}}}\frac{\Omega }{c}}{\sqrt{%
g_{00}+2g_{0\phi }\frac{\Omega }{c}+g_{\phi \phi }\frac{\Omega ^{2}}{c^{2}}}}
\label{tofomega}
\end{equation}

Given the conditions we have chosen, $\Omega $ is of course not a free parameter.
Since the metric tensor does not depend on $\phi $ and $t$ we have two
constants of motion: $E$ (energy per unit mass) and $L$ (angular momentum
per unit mass). The time and $\phi $ components of the four-velocity of a
test mass, $u$, are:

\[
\left\{
\begin{array}{c}
u^{0}=\frac{Eg_{\phi \phi }-Lg_{0\phi }}{g_{00}g_{\phi \phi }-g_{0\phi }^{2}}
\\
u^{\phi }=\frac{Lg_{00}-Eg_{0\phi }}{g_{00}g_{\phi \phi }-g_{0\phi }^{2}}%
\end{array}%
\right.
\]
Then we obtain, for any free fall, including circular orbits:

\begin{equation}
\frac{\Omega }{c}=\frac{u^{\phi }}{u^{0}}=\frac{d\phi }{cdt}=\frac{%
Lg_{00}-Eg_{0\phi }}{Eg_{\phi \phi }-Lg_{0\phi }}  \label{omega}
\end{equation}

Introducing (\ref{omega}) into (\ref{tofomega}) we get:

\begin{equation}
\Delta T_{\Omega }=\frac{4\pi }{c}\frac{g_{0\phi }}{\left( g_{00}\right)
^{3/2}}E\sqrt{\frac{g_{00}g_{\phi \phi }-g_{0\phi }^{2}}{g_{00}L^{2}-2g_{0%
\phi }LE+g_{\phi \phi }E^{2}}}  \label{deltom}
\end{equation}

%------------------------Section-------------------------
\section{Kerr-Newman space-time}
%------------------------Section-------------------------

As stated in the Introduction, we start from the empty space-time
surrounding a spinning mass. The number of exact solutions of the Einstein
equations is quite small, but hopefully we have one precisely in the case of
a spinning mass; specifically the central object turns out to be a rotating
black hole with a quasi-spherical event horizon. We are of course referring
to the Kerr solution \cite{kerr}.

In order to be a bit more general we shall include the possibility that the
central source be also electrically charged: this is known as the
Kerr-Newman \cite{newman1965metric,wald2010general} solution.\footnote{%
For completeness we must specify that the Kerr-Newman solution does not
include in the Einstein equations a cosmological constant: in practice, it
does not consider the presence of \textit{dark energy. }Here, however, we
are not concerned with the cosmic scale.} The properties of such a
space-time are synthesized in the line element:

\begin{eqnarray}
ds^{2} &=&\frac{r^{2}-2mr+a^{2}+q^{2}-a^{2}\sin ^{2}\theta }{r^{2}+a^{2}\cos^{2}
\theta }c^{2}dt^{2}+2a\sin ^{2}\theta \left( \frac{2mr-q^{2}}{%
r^{2}+a^{2}\cos^{2} \theta }\right) cdtd\phi   \nonumber \\
&&-\frac{\left( r^{2}+a^{2}\right) ^{2}-\left( r^{2}-2mr+a^{2}+q^{2}\right)
a^{2}\sin ^{2}\theta }{r^{2}+a^{2}\cos^{2} \theta }\sin ^{2}\theta d\phi ^{2}
\label{KN} \\
&&-\frac{r^{2}+a^{2}\cos^{2} \theta }{r^{2}-2mr+a^{2}+q^{2}}dr^{2}-\left(
r^{2}+a^{2}\cos^{2} \theta \right) d\theta ^{2}  \nonumber
\end{eqnarray}

Here we use Boyer-Lindquist coordinates. The meaning of the physical
parameters (all lengths) is summarized as follows:

\begin{equation}
m=G\frac{M}{c^{2}};\text{ }a=\frac{J}{Mc};\text{ }q^{2}=\frac{GQ^{2}}{4\pi
\varepsilon _{0}c^{4}}
\end{equation}

$M$ is interpreted as the mass of the source, $J$ its angular momentum, $Q$
its charge; $G$ is the Newton constant and $\varepsilon _{0}$ is the
dielectric constant of the vacuum.

In order to evidence the travel time asymmetry in a Sagnac-like configuration we may use any "messenger"
travelling along a closed path (in space). The simplest is to refer to light
so that it is $ds=0$.

The standard result is the one shown in Eq. (\ref{eq:formulafond}) which, for a
circular trajectory in the equatorial plane, coincides with Eq. (\ref{eq:fondcirc}).

The sign of the tof difference depends on which sense we assume to be positive, so it is
irrelevant provided the next steps be consistent with the initial choice.
For our purposes and without reduction of generality we may choose a closed
circuit in the equatorial plane ($\theta =\pi /2$) and at a constant radius (%
$r=R$), i.e. a circle. Under these conditions formula (\ref{eq:fondcirc}) holds and the result is:

\begin{equation}
\Delta T=\frac{4\pi }{c}\frac{g_{0\phi }}{g_{00}}=\frac{4\pi }{c}a\frac{%
2mR-q^{2}}{R^{2}-2mR+q^{2}}  \label{tof}
\end{equation}

Coordinate time coincides with the time of a distant observer at rest with
the mass in the origin. Space-time surrounding such an observer is flat: we
may verify it letting $r$ go to $\infty $ in the line element of Eq. (\ref{KN})%
. The metric tensor there is asymptotically diagonal, so, should the closed
path of light be entirely located in those far regions ($R\rightarrow \infty
$), the tof asymmetry would also tend to zero: result emerging from Eq. (\ref{tof}).

Still using coordinate time, but getting closer and closer to the central
mass we may see what happens at the other end of the radial distances, i.e.
for $R\rightarrow 0$. Remarkably the tof asymmetry is finite and non-zero:

\begin{equation}
\left. \Delta T\right\vert _{R=0}=\frac{4\pi }{c}a=4\pi \frac{J}{Mc^{2}}
\label{tof0}
\end{equation}

By the way, if the closed path revolves around the axis of the central mass,
but is not contained in the equatorial plane, the result is different.
Considering a circular circuit in the plane $r\cos \theta =z_{0}$ constant,
the time asymmetry is

\begin{equation}
\left. \Delta T\right\vert _{R\cos \Theta =z_{0}}=-\frac{4\pi }{c}\frac{%
a\left( 2mR-q^{2}\right) \sin ^{2}\Theta }{R^{2}-2mR+a^{2}+q^{2}-a^{2}\sin
^{2}\Theta }  \label{tofa}
\end{equation}

Now $\Theta $ is the half-aperture of a cone having its vertex in the
origin. Introducing the variable $\rho =R\sin \Theta $ such that $R=\sqrt{%
\rho ^{2}+z_{0}^{2}}$. (\ref{tofa}) becomes

\[
\left. \Delta T\right\vert _{R\cos \Theta =z_{0}}=-\frac{4\pi }{c}\frac{%
a\left( 2m\sqrt{\rho ^{2}+z_{0}^{2}}-q^{2}\right) \frac{\rho ^{2}}{\rho
^{2}+z_{0}^{2}}}{\rho ^{2}+z_{0}^{2}-2m\sqrt{\rho ^{2}+z_{0}^{2}}%
+a^{2}+q^{2}-a^{2}\frac{\rho ^{2}}{\rho ^{2}+z_{0}^{2}}}
\]%
and letting $\rho $ go to $0$, i.e. reducing the circuit to the axis above
or under the origin, we see that

\[
\left. \Delta T\right\vert _{\rho \rightarrow 0}=0
\]%
The asymmetry disappears.

It is only in the origin, where the source is located, that the asymmetry is
the non-zero (\ref{tof0}). Being at a point, this difference may be read as
a time uncertainty there.

We may think to apply this result to an electron, so that the eigenvalue of
the axial component of the angular momentum (which is the semiclassical
interpretation of the spin) is $J=\hbar /2$. Eq. (\ref{tof0}) now reads:

\begin{equation}
\Delta T=\frac{h}{Mc^{2}}  \label{HeiKer}
\end{equation}

Considering that $Mc^{2}$ is the relativistic (rest) energy of the particle,
(\ref{HeiKer}) coincides with the lower limit of the Heisenberg uncertainty
formula. Remarkably, the result does not depend neither on $G$ nor on $q$:
the charge looses any role and even the coupling of the mass to the
curvature of space-time disappears. What matters is not the curvature but
only the chiral symmetry of space-time induced by the rotation of the mass.

We have made reference to an electron, both because it is a Fermion and
because it is a lepton, then has no internal structure; but, should we have
considered a proton the result would have been the same, and, apart for a
factor of the order of $2$, also bosons would have been OK.

The result we have found holds for a distant observer, not directly affected by the curvature and peculiar symmetry of space-time. If instead we look at a local observer orbiting the central mass and getting closer and closer to it, the formula to be used is (\ref{deltom}) and for this observer the time of flight asymmetry in the origin falls to zero. The details of this calculation are developed in the Appendix.

\section{Concluding remarks}

Our aim, in this paper, was simply to evidence an unexpected link between
classical GR formulae and a fundamental relation of quantum mechanics as the
Heisenberg uncertainty principle is. As we have seen, the mentioned link
appears when considering the asymmetry in the times of flight of light beams
coming back to the observer after travelling in opposite directions along
the same closed contour (actually the result holds also for other
messengers, provided their local velocity along the trajectory depends only
on the position and not on the rotation sense). The interesting result
emerges when the space-time of a steadily spinning mass is considered and
the closed path shrinks to the origin where the central source of gravity is
located. In order to be able to call in the quantum spin of a particle we
have used an explicit and exact solution of the Einstein equations, the
Kerr-Newman solution, assuming that the central source be an elementary
Fermion (a lepton, like an electron, without more internal degrees of
freedom). Under these conditions the difference between the times of flight
in different rotation senses reduce to an uncertainty on the time measured
at the origin;\ identifying the classical angular momentum of the source
with the spin of the Fermion and the central mass with the rest mass of the
particle, coinciding with its energy modulo $c^{2}$, we have obtained the
lowest value of the Heisenberg formula for energy and time uncertainties.
The result does not depend neither on the constant of gravity $G$ nor on the
charge of the particle, so we may interpret it as a property of the symmetry
of space-time when it couples to elementary spinning sources.

We think this result may be inspiring for further work, requiring deeper and
more detailed elaborations leading along a path that promises more surprises.

\appendix
\section{Asymmetry for orbiting observers and Clock Effect}

For completeness we must recall that the calculation leading to Eq. (\ref{HeiKer}) hold for coordinate time
and a non-rotating observer, i.e. for an inertial observer located far away.
If we look at the situation from the viewpoint of an observer orbiting the
central mass at a finite distance from the origin and in the equatorial
plane, the formula to be used is (\ref{deltom}), i.e. explicitly

\begin{equation}
\Delta T_{\Omega }=\frac{4\pi }{c}E\frac{aR^{2}\left( 2mR-q^{2}\right) }{%
\left( q^{2}+R^{2}-2mR\right) ^{\frac{3}{2}}}\sqrt{\frac{%
-a^{2}-q^{2}-R^{2}+2mR}{\left( R^{2}-2mR+q^{2}\right) L^{2}-2\left(
2mR-q^{2}\right) aLE-\left( 2ma^{2}R+R^{4}-a^{2}q^{2}+a^{2}R^{2}\right) E^{2}%
}}  \label{rot}
\end{equation}

The result now tends to $0$ both for $R\rightarrow 0$ and for $R\rightarrow
\infty $.

Considering the clock effect, the basic formula comes
from (\ref{TC}) and is:

\begin{equation}
\Delta T_{C}=\frac{2\pi }{\omega _{+}-\omega _{-}}\left( \sqrt{%
g_{00}+2g_{0\phi }\frac{\omega _{+}}{c}+g_{\phi \phi }\frac{\omega _{+}^{2}}{%
c^{2}}}-\sqrt{g_{00}+2g_{0\phi }\frac{\omega _{-}}{c}+g_{\phi \phi }\frac{%
\omega _{-}^{2}}{c^{2}}}\right)   \label{deltaTC}
\end{equation}%
where now $\omega _{+}$ and $\omega _{-}$ are the free fall angular
velocities in opposite directions along the same orbit. Let us say that one
corresponds to $L=L_{+}>0$ and the other to $L_{-}=-L_{+}<0$. Recalling Eq. %
(\ref{omega}) it is

\[
\left\{
\begin{array}{c}
\frac{\omega _{+}}{c}=\frac{Lg_{00}-Eg_{0\phi }}{Eg_{\phi \phi }-Lg_{0\phi }}
\\
\frac{\omega _{-}}{c}=-\frac{Lg_{00}+Eg_{0\phi }}{Eg_{\phi \phi }+Lg_{0\phi }%
}%
\end{array}%
\right.
\]

Let us introduce these value into Eq. (\ref{deltaTC}):

\[
\Delta T_{C}=\frac{\pi }{c}\frac{\left( g_{\phi \phi
}^{2}E^{2}-L^{2}g_{0\phi }^{2}\right) }{EL\left( Lg_{0\phi }-Eg_{\phi \phi
}\right) }\frac{\sqrt{\allowbreak 2g_{0\phi }LE-g_{00}L^{2}-g_{\phi \phi
}E^{2}}-\sqrt{\allowbreak -\left( g_{00}L^{2}+2g_{0\phi }LE+g_{\phi \phi
}E^{2}\right) }}{\sqrt{\left( g_{0\phi }^{2}-g_{00}g_{\phi \phi }\right) }}
\]

which in the case of a Kerr-Newman space-time becomes

\footnotesize
\[
\Delta T_{C}=\frac{\pi }{cEL}\frac{\left(
ER^{4}+Ea^{2}R^{2}+2mEa^{2}R-2LmaR+Laq^{2}-Ea^{2}q^{2}\right) }{R^{3}}\frac{%
F_{1}-F_{2}}{%
\sqrt{a^{2}+q^{2}+R^{2}-2mR}}
\]
\normalsize
where
\[
F_{1}=\sqrt{\allowbreak 2\left( 2mR-q^{2}\right) aLE-\left(
R^{2}-2mR^{2}+q^{2}\right) L^{2}+\left( \allowbreak
R^{4}+a^{2}R^{2}+2ma^{2}R-a^{2}q^{2}\right) E^{2}}
\]
and
\[
F_{2}=\sqrt{\allowbreak -\left(
\left( R^{2}-2mR^{2}+q^{2}\right) L^{2}+2\left( 2mR-q^{2}\right) aLE-\left(
\allowbreak R^{4}+a^{2}R^{2}+2ma^{2}R-a^{2}q^{2}\right) E^{2}\right) }
\]

$\allowbreak $When two counter-orbiting identical clocks are compared in the
same position along their circular and common  orbit they appear to have
lost synchrony by the amount $\Delta T_{C}$ per turn. The asynchrony
disappears when $a=0$, i.e. when the source of gravity is not spinning.

We also see that $\Delta T_{C}$ becomes $0$ at the horizon $R=q^{2}/2m$ and
is of course $0$ for $R\rightarrow \infty $.

%\bibliography{editorialML.bib}

\end{document}